\documentclass[aps,pre,twocolumn,showpacs,superscriptaddress,groupedaddress,reprint]{revtex4-1}
\usepackage{subfig}
\usepackage{wrapfig}
\usepackage{textcomp}
\usepackage{inputenc}
\usepackage{enumitem}
\usepackage{graphicx}
\usepackage{amssymb}
\usepackage{amsmath}
\usepackage{bm}
\usepackage{amsmath, amsthm, amssymb}
\usepackage{hyperref}
\usepackage{color}
\usepackage{rotating}
\usepackage{afterpage}
\usepackage{tabulary,booktabs}
\usepackage{amsfonts}
\usepackage{dcolumn}
\usepackage{float}
\usepackage{bm}
\usepackage{natbib}
\usepackage{caption}
\usepackage{verbatim}

\begin{document}
\setcounter{table}{0}

\title{Sommerfeld expansion of electronic entropy \\ in {\sc INFERNO}-like average atom model}

\author{Philippe Arnault}
\email{philippe.arnault@cea.fr}
\affiliation{CEA, DAM, DIF, F-91297 Arpajon, France}
\author{Julien Racine}
\affiliation{CEA, DAM, DIF, F-91297 Arpajon, France}
\author{Augustin Blanchet}
\affiliation{CEA, DAM, DIF, F-91297 Arpajon, France}
\affiliation{Universit\'e Paris-Saclay, CEA, Laboratoire Mati\`ere en Conditions Extr\^emes, 91680 Bruy\`eres-le-Ch\^atel, France}
\author{Jean-Pierre Raucourt}
\affiliation{CEA, DAM, DIF, F-91297 Arpajon, France}
\author{Jean-Christophe Pain}
\affiliation{CEA, DAM, DIF, F-91297 Arpajon, France}
\affiliation{Universit\'e Paris-Saclay, CEA, Laboratoire Mati\`ere en Conditions Extr\^emes, 91680 Bruy\`eres-le-Ch\^atel, France}
\date{\today}

\begin{abstract}

In average atom (AA) model, the entropy provides a route to compute thermal electronic contributions to the equation of state (EOS). The complete EOS comprises in many modelings an additional 0K-isotherm and a thermal ionic part. Even at low temperature, the AA model is believed to be the best practical approach. However, when it comes to determine the thermal electronic EOS at low temperatures, the numerical implementation of AA models faces convergence issues related to the pressure ionization of bound states. At contrast, the Sommerfeld expansion tells us that the variations with temperature of thermodynamic variables should express in simple terms at these low temperatures. This led us to tackle the AA predictions with respect to the Sommerfeld expansion of the electronic entropy. We performed a comprehensive investigation for various chemical elements belonging to $s-$, $p-$, $d-$, and $f-$blocks of the periodic table, at varying densities. This was realized using an {\sc Inferno}-like model since this approach provides the best theoretical framework to address these issues. Practical prescriptions are provided as functions of the atomic number.
\end{abstract}

\maketitle

\section{Introduction}

Equation of state (EOS) relates the state variables of a material: pressure, temperature, internal energy, and so on. It also delineates the frontiers of the solid phases, the liquid, gas, and plasma ones \cite{Wallace2003}. For these reasons, EOS contributes to the foundation of various process engineerings, technological applications, and astrophysics. Furthermore, the study of matter under extreme conditions is crucial for the understanding of the physics of stellar and planetary interiors as well as for the design and interpretation of inertial confinement fusion (ICF) experiments \cite{Drake2018}. The ICF modeling consists in hydrodynamic simulations requiring reliable EOS data for a variety of elements over a wide range of material conditions, from the solid state to the rarefied plasma and high energy-density regimes \cite{Gaffney2018}.
The large range of conditions and their extreme nature make it impossible to completely explore the EOS experimentally and thus theoretical modelings, benchmarked against experiments at selected density-temperature conditions, are determinant \cite{Fortov1998}. The most accurate theoretical approaches consist in \textit{ab initio} simulations within density functional theory (DFT) \cite{Jones2015, Yu2016} or path integral Monte Carlo formalism \cite{Bonitz2020}. Unfortunately, these techniques are so computationally demanding that they are often viewed as scarce, albeit essential, as well as experiments \cite{Blanchet2020}. Since the hydrodynamics requires stringent constraints of conservation of mass, momentum, and energy \cite{arnault2022}, EOS approximations come often along with a free energy model, which warrants the thermodynamic consistency. In this context, the average atom (AA) model has become to be instrumental (see \cite{Callow2022} and references therein).

EOS models typically consist of three parts: the cold curve, the ion-thermal contribution and the electron-thermal contribution \cite{Chisolm2003}. This separation aims at taking the best from the theoretical modelings for each contribution. AA models are used for the electron-thermal part \cite{Swift2019} whereas the cold curve benefits from experimental measurements and is guided by solid-state physics \cite{Vinet1989}.
The ion-thermal contribution takes into account experiments at low temperature, but must also reproduces the wide-range behavior from the solid, the liquid up to more or less correlated plasmas \cite{Kerley1980, Johnson1991, Swift2020}.

In this paper, we focus on the AA predictions in the low temperature regime, characteristic of solids and liquid metals. Since the electron-thermal contribution can be small in this regime \cite{Wallace2003}, the experiments are often dominated by the other contributions. Therefore, the modeler is left alone with theoretical tools. A quick look at the literature about the multi-phase EOS shows that there is no consensus as to the best way to account for the electron-thermal contribution. Simple prescriptions rest on the free electron gas (FEG) model using the Thomas-Fermi (TF) ionization \cite{Bhattacharya2014} or some Sommerfeld-like approximation where thermodynamic quantities are expressed as power laws in temperature and density \cite{Benedict2014}. More involved prescriptions use the Thomas-Fermi (TF) average atom model \cite{Thomas1927, Fermi1927, Feynman1949, Benedict2014, Cochrane2022} or its quantum-mechanical extensions \cite{Benedict2014, Wu2021}, before resorting to \textit{ab initio} simulations \cite{Meyer-ter-Vehn1988}. Although the AA model is considered as the most realistic of the practical approaches, 
it faces convergence issues at low temperature when pressure ionization of fully occupied bound state occurs \cite{More1985}. This is to be contrasted with the simplicity of the Sommerfeld expansion of thermodynamic variables in powers of the temperature \cite{Sommerfeld1928, Ashcroft1976, Wetta2019}. 

Here, we contribute to the modeling of the electron-thermal EOS at low temperature using an {\sc Inferno}-like AA model \cite{Liberman1979, Wilson2006, Penicaud2009a} to compare its predictions with the Sommerfeld approximation. This kind of AA model, which derives variationally from a free energy functional, is recognized as one of the best approaches to tackle the pressure ionization of bound states into resonances, which extend the representation of localized charge from the spectrum of bound states to the free energy continuum \cite{More1985}. This avoids the appearance of discontinuities of thermodynamic variables and allows us to compare them with their Sommerfeld expansion. We extensively studied four elements: hydrogen, aluminum, iron, and cerium, which belong to $s-$, $p-$, $d-$, and $f-$blocks of the periodic table. For selected conditions, we also include a larger range in atomic number $Z$. Furthermore, we chose to work with the electronic entropy since it easily provides the thermal EOS as is explained in App.\,\ref{EOS} and is straightforwardly related to the density of states (DOS).

In a first theoretical part, we briefly recall  the tenets of the AA modeling in Sec.\,\ref{INFERNO} in order to set the problematics of pressure ionization and to introduce the relevant quantities. We also go into the details of the Sommerfeld approximation in Sec.\,\ref{Sommerfeld} and present a direct Sommerfeld-like approach to the electronic entropy in Sec.\,\ref{Sommerfeld_S}, which has not been presented so far in the literature to our knowledge. This theoretical part is ended in Sec.\,\ref{FEG} by a presentation of the FEG model as a useful concept in the present context.

The results of an {\sc Inferno}-like model are described in Sec.\,\ref{rst} for hydrogen (Sec.\,\ref{H}), aluminum (Sec.\ref{Al}), iron (Sec.\,\ref{Fe}), and cerium (Sec.\,\ref{Ce}) in a large range of density from 10$^{-4}$ to 10$^4$ g.cm$^{-3}$ at low temperature from 0.01 to 10\,eV. They are further analyzed in Sec.\,\ref{model} where the domain of validity of the Sommerfeld approximation is delineated (Sec.\,\ref{Sommerfeld_T}). This analysis allowed us to extend the study to a larger set of elements in order to evidence any trends with the atomic number $Z$ (Sec.\,\ref{Z}). In particular, the FEG model is shown to be adequate for very dense system only (Sec.\,\ref{powerlaw}). Finally, the sensitivity of the results to the exchange and correlation (XC) functional is addressed in Sec.\,\ref{xc}.

\section{Theory}
\label{theory}

\subsection{{\sc INFERNO}-like average atom}
\label{INFERNO}

The principle behind the average atom modeling is to represent the many-body system of electrons and nuclei by an equivalent unique atom submitted to appropriate boundary conditions \cite{Callow2022}. The most obvious constraint is to confine the average atom is a volume $V$ related to the mass density $\rho$ of the medium. Many modelings differ in the way this confinement is realized \cite{Liberman1979, Crowley2001, Perrot1995a, Starrett2013, Starrett2020, Blenski2023}. The most simple one, used in {\sc Inferno}, involves the Wigner-Seitz (WS) sphere of radius $r_\text{ws}$ given by
\begin{equation}
\frac{4}{3} \pi r_\text{ws}^3 \, n_i = 1,
\end{equation}
where $n_i = \rho \, {N_A}/{A}$ is the number density, $A$ the molar mass of the chemical element, and $N_A$ the Avogadro number. The constraint states that the volume $V = 1/n_i$ containing the average atom should be neutral.

The electronic structure of this atom is computed within the framework of DFT \cite{Hohenberg1964, Kohn1965, Mermin1965, Parr1989}. This leads, in particular, to an average atom with fractional occupancy $n_s$ of the bound states $\varepsilon_s$ equal to the product of their degeneracy $g_s$ by the Fermi-Dirac (FD) function $f(\varepsilon_s)$
\begin{equation}
n_s = g_s\,f(\varepsilon_s),
\end{equation}
where $s$ stands for the principal quantum number $n$ and the quantum numbers  $j$, $\ell$, and $m$ are associated respectively with the total angular momentum $J$, the orbital angular momentum $L$, and its projection $L_z$. The same is true for free states except that the index $s$ stands for the energy $\varepsilon$, or the wave-vector $k$, instead of the principal quantum number $n$.

The {\sc Inferno} model was the first to propose a consistent framework where the equations were obtained through the minimization of a free-energy functional \cite{Liberman1979, Wilson2006, Penicaud2009a}. This warrants \textit{a priori} the thermodynamic consistency of the results. The model considers an atom as a point nucleus surrounded by its $Z$ electrons and places it at the center of the WS spherical cavity buried in a jellium, \emph{i.e.}, a uniform distribution of positive charges which takes place of the surrounding ions, and a constant electron distribution that ensures electrical neutrality.
Electronic structure is then computed in a self-consistent way, where the effective potential $V_\text{eff}(\vec r)$ within the cavity is a functional of the electron density $n_e(\vec r)$ itself computed from the wave-functions subject to the effective potential. Outside the cavity, the latter is assumed to be constant.

Every electron is represented by a wave-function $\psi_s(\vec r)$ that is a solution of the Dirac equation, with exponentially decreasing boundary condition at large distance for bound states and oscillating behavior for free states. 
In the WS sphere, the effective potential $V_\text{eff}(r)$ is spherically symmetric and the wave function $\psi_s$ is composed of two radial components, the major component $F_s(r)$ and the minor one $G_s(r)$, at contrast to the radial Schrodinger equation that only needs one. These radial components are solutions to the following system of equations \cite{Nikiforov2005}
\begin{eqnarray}
&\dfrac{dF_s}{dr} &= - \dfrac{\kappa_s}{r}\,F_s(r) - \dfrac{1}{c}\left( V_\text{eff}(r) - c^2 - \varepsilon_s \right)\,G_s(r),  \nonumber \\  \\
&\dfrac{dG_s}{dr} &= + \dfrac{\kappa_s}{r}\,G_s(r) + \dfrac{1}{c}\left( V_\text{eff}(r) + c^2 - \varepsilon_s \right)\,F_s(r), \nonumber
\end{eqnarray}
where atomic units (a.\,u.) $e = \hbar = m_e = 1$ are
used and we choose to keep the speed of light $c$ instead of the hyperfine structure constant $\alpha$. The orbital parameter $\kappa_s$ is
\begin{equation}
 \kappa_s = \left\{
    \begin{array}{ll}
        -(\ell+1) & \mbox{if } j = \ell+ \frac{1}{2}, 
        \\ \\
        +\,\ell & \mbox{if } j = \ell - \frac{1}{2}.
    \end{array}
\right.
\end{equation}
Inside the cavity, the effective potential $V_\text{eff}$ comprises the interaction with the nucleus of charge $Z$, the electrostatic interaction with the total electron density $n_e$, and the XC potential $V_\text{xc}$
\begin{equation}
V_\text{eff}(r) = - \dfrac{Z}{r} + \int_{r' \le r_\text{ws}} \dfrac{n_e(r')}{|\vec r - \vec r\,'|} \, d^3\vec r\,' + V_\text{xc}(r).
\end{equation}
One of the numerical challenges in the solution of these equations concerns the dissolution of bound states within the continuum of free states as the density increases. This phenomenon, known as "pressure ionization", leads to particular issues at low temperature when the bound states are completely occupied since the large localized density associated with the bound state cannot abruptly transit to a situation of complete delocalization. Instead, the solution exhibits sharp resonances in the continuum that are the scars of the disappearing bound states. As the compression goes on increasing, these resonances enlarge and progressively dissolve into the continuum. The detection and the discretization of these resonances represent a cumbersome numerical challenge.

The electronic structure (bound and free states) being known, the thermodynamic quantities can be calculated. A theoretical issue then shows up that is best explained by Liberman \cite{Liberman1979}: \textit{"In devising this model of condensed matter, where a single atom is embedded in an electron gas, we introduced the electron gas to simulate the effects of the surrounding atoms on a particular one. However, it is the properties of the atom -- its energy, for example -- which concern us. However, the mathematical expressions for these quantities include large contributions from the electron gas."} To circumvent this issue, Liberman proposed two main versions of {\sc Inferno}: the so-called A and T models, differing by the calculation of thermodynamic quantities. In the T model, the quantities related to the jellium are subtracted from the quantities calculated in the whole space, whereas in the A model, the separation between the ionic cell and the jellium is of spatial nature. For instance, the {\sc Purgatorio} code \cite{Wilson2006} is very close to the A model. Liberman's model-A states that thermodynamic quantities must be either obtained from integrations limited to the WS confinement sphere or multiplied by the factors $X(\varepsilon_s)$ giving the fraction of the density, associated with the wave-function $\psi_s$, pertaining to the confinement volume
\begin{eqnarray}
&X(\varepsilon_s)\, &= \int_{r \le r_\text{ws}} \psi_s(\vec r)\, \psi_s^*(\vec r)\, d^3 r, \\
& &= 2 \,|\kappa_s| \int_0^{r_\text{ws}} \left[F_s^2(r) +G_s^2(r)\right] \, dr. \nonumber
\end{eqnarray}
This model-dependent prescription represents a step outside a thermodynamic consistent framework, but this is the price to connect quantities obtained from the average atom to quantities representative of the whole system. It is expected that inconsistency may appear only when $f(\varepsilon_s)$ is close to 1 and $X_s$ far less than 1. Unfortunately, this happens just during the transition from bound states toward sharp resonances. Since the model is precisely built to address this kind of issues, it is only \textit{a posteriori} that a consistency check can be performed.

In the following, we shall need to consider the density of states $n(\varepsilon)$ and the entropy $S$ per atom given respectively by
\begin{equation}
\label{eq:DOS_AA}
n(\varepsilon) = \dfrac{1}{V}\,\left[\, \sum_{s \text{\,bound}} \delta(\varepsilon-\varepsilon_s) + H(\varepsilon-V_\infty) \right] \,  X(\varepsilon),
\end{equation}
where $H$ is the Heaviside function, $\delta$ the Dirac distribution, and
\begin{equation}
\label{eq:S}
\dfrac{S}{V} = - k_B \int_{-\infty}^{+\infty}  n(\varepsilon) \, s(\varepsilon)\,d\varepsilon  + \dfrac{S_\text{xc}}{V},
\end{equation}
where
\begin{equation}
s(\varepsilon) = f(\varepsilon) \,\log f(\varepsilon) + [1-f(\varepsilon)]\, \log[1- f(\varepsilon)],
\end{equation}
and $S_\text{xc}$ comes from the XC free energy functional. This contribution to the exchange and correlation is important in partially degenerate plasmas, but can be safely neglected at the low temperatures where the Sommerfeld expansion is valid \cite{Karasiev2014}. As we shall see in Sec.\,\ref{model}, the validity domain of Sommerfeld's approximation has an upper limit in the degeneracy parameter $\theta$ of the order of 0.05.

It is instructive to discuss the behavior of the function $s(\varepsilon)$ at low temperature. As Fig.\,\ref{fig:FD} shows, it vanishes wherever $\varepsilon$ differs from the chemical potential $\mu$ by a few $k_BT$. Therefore, the entropy $S$ is built upon from the values of the density of states $n(\varepsilon)$ in the vicinity of the chemical potential $\mu$. At the same time, the value of the chemical potential $\mu$ is determined from the neutrality of the WS confinement sphere. Let us recall the definition of $n(\varepsilon)$ in Eq.\,\eqref{eq:DOS}
\begin{equation}
\label{eq:AA_Z}
\dfrac{Z}{V} = \int_{-\infty}^\infty n(\varepsilon) f(\varepsilon)\, d\varepsilon.
\end{equation}

\subsection{Sommerfeld expansion}
\label{Sommerfeld}

\begin{figure}[!t]
\begin{center}\includegraphics[width=0.45\textwidth]{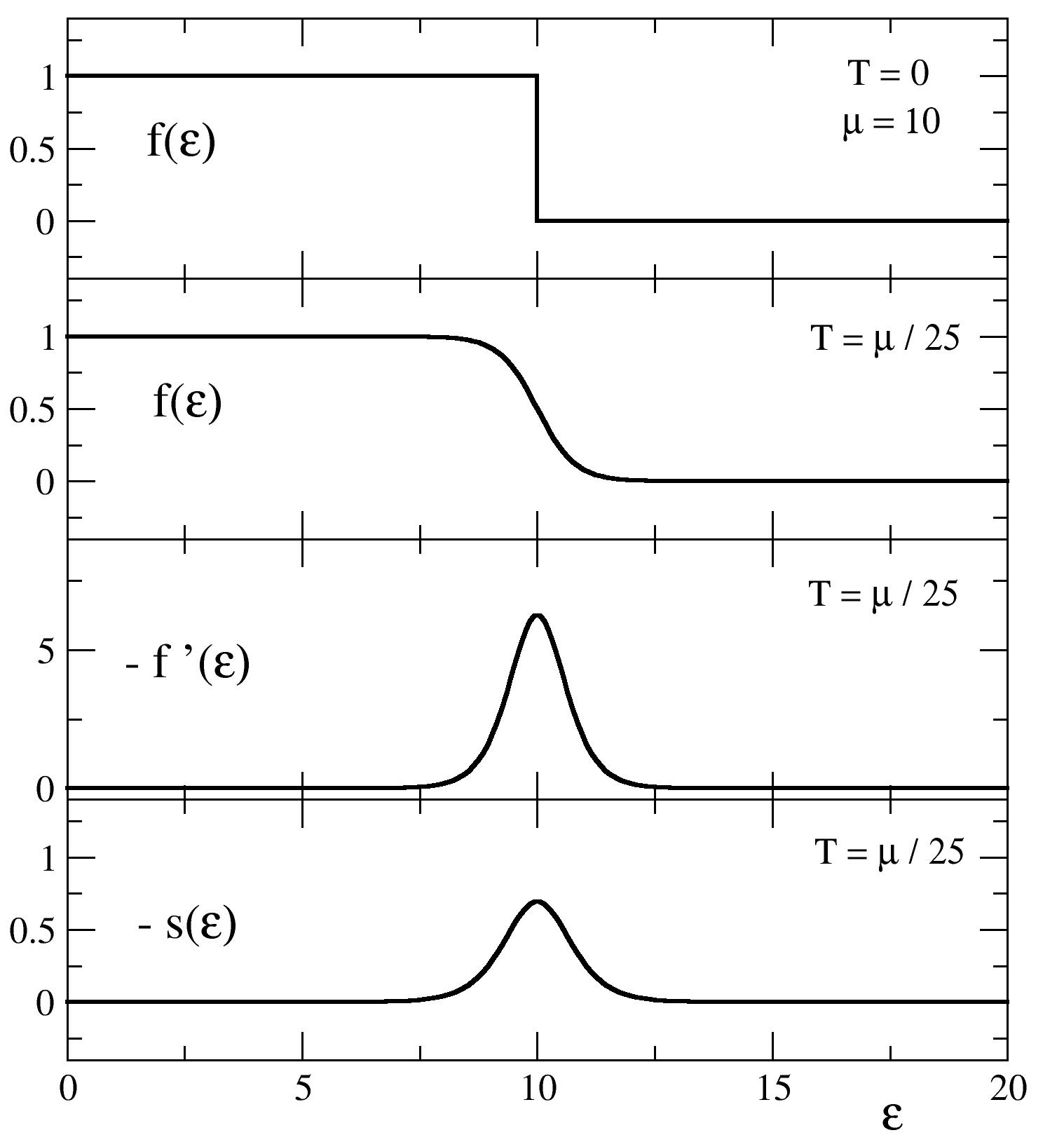}
\caption{Behavior of the Fermi-Dirac distribution $f(\varepsilon)$ at low temperature. For the sake of clarity, we choose a  temperature  $T = \mu/25$ close to the highest temperatures where the Sommerfeld expansion is valid. Also shown are the derivative of $f(\varepsilon)$ and the entropy factor $s(\varepsilon)$. They are sharply peaked and symmetric around $\mu$.}
\label{fig:FD}
\end{center}
\end{figure}

Kohn--Sham (KS) density functional theory (DFT) describes an equilibrium situation, where the electrons can be treated as independent particles moving in an effective potential $V_\text{eff}(\vec r)$ that depends on their density $n_e(\vec r)$. At finite temperature $T$ \cite{Mermin1965}, the energy levels $\varepsilon$ of electrons are occupied according to the FD statistics $f(\varepsilon)$ which involves the inverse temperature $\beta = 1/k_B T$ and the chemical potential $\mu$ \cite{Ashcroft1976}
\begin{equation}
f(\varepsilon) = \dfrac{1}{e^{\beta (\varepsilon-\mu)} + 1}.
\end{equation}
Therefore, the density $n_e$, as any functional of it, depends  explicitly on temperature through the FD function $f(\varepsilon)$. This function is a Heaviside $H(\varepsilon_F -\varepsilon)$ at $T = 0$, transiting from 1 to 0 at the Fermi energy $\varepsilon_F$, which is defined as the highest energy level occupied by electrons at $T = 0$. As the temperature $T$ increases from 0, the step-like jump is located at the value of the chemical potential $\mu$ and smooths out over a range of $k_B T$ around $\mu$, as illustrated in Fig.\,\ref{fig:FD}. At high enough temperature, the FD function  $f(\varepsilon)$ tends to the Maxwell--Boltzmann distribution.

Sommerfeld expansion \cite{Sommerfeld1928, Ashcroft1976} is a low-$T$ Taylor series of integrals of the form
\begin{equation}
I(T,\mu) = \int_{-\infty}^\infty g(\varepsilon) f(\varepsilon)\, d\varepsilon,
\end{equation}
where $g(\varepsilon)$ is a function vanishing  as $\varepsilon \to -\infty$ and diverging no more rapidly than some power of $\varepsilon$ as $\varepsilon \to +\infty$. Introducing the primitive $G(\varepsilon)$
\begin{equation}
G(\varepsilon) = \int_{-\infty}^\varepsilon g(\varepsilon') \, d\varepsilon',
\end{equation}
an integration by part gives
\begin{equation}
I(T,\mu) = - \int_{-\infty}^\infty G(\varepsilon) \,f'(\varepsilon)\, d\varepsilon,
\end{equation}
with
\begin{equation}
-f'(\varepsilon) = -\frac{\partial f}{\partial \epsilon} = \beta \, f(\varepsilon) \left[1 - f(\varepsilon)\right].
\end{equation}
As depicted in Fig.\,\ref{fig:FD}, the derivative of the FD distribution $f(\varepsilon)$ is sharply peaked at the value of the chemical potential $\mu$ at low temperature ($\theta = k_B T/\varepsilon_F \ll 1$). The integral $I(T,\mu)$ and its approximations are therefore highly sensitive to the Taylor expansion of $G(\epsilon)$ around $\varepsilon = \mu$. Provided that this function  is not too rapidly varying in this neighborhood, just a few terms should be needed. Since the derivative $f'(\varepsilon)$ is symmetric around the value of the chemical potential $\mu$, the first non-vanishing contribution to  $I(T,\mu)$, in addition to $G(\mu)$, is of second order in the Taylor expansion of $G$ 
\begin{equation}
G(\varepsilon) = G(\mu) + (\varepsilon-\mu)\, G'(\mu) + \frac{1}{2}\,(\varepsilon-\mu)^2 \, G''(\mu).
\end{equation}
At second order, one obtains
\begin{equation}
I(T,\mu) = G(\mu) - \frac{1}{2}\, G''(\mu)\, \int_{-\infty}^\infty (\varepsilon-\mu)^2 \,f'(\varepsilon)\, d\varepsilon.
\end{equation}
To evaluate the remaining integral, the following change of variable is done
\begin{equation}
z = \beta (\varepsilon - \mu),
\end{equation}
leading to
\begin{equation}
I(T,\mu) = G(\mu) + a_1\, G''(\mu)\,(k_B T)^2,
\end{equation}
with 
\begin{equation}
a_1 = \frac{1}{2}\, \int_{-\infty}^\infty z^2\,\dfrac{e^z}{(1+e^z)^2}\,dz.
\end{equation}
Using the Taylor expansion at all orders leads to the Sommerfeld expansion \cite{Sommerfeld1928,Ashcroft1976}:
\begin{equation}
\int_{-\infty}^\infty g(\varepsilon) f(\varepsilon)\, d\varepsilon = \int_{-\infty}^\mu g(\varepsilon) \, d\varepsilon + \sum_{n=1}^\infty a_n\,(k_B T)^{2n}\,\dfrac{d^{2 n-1} g}{d\varepsilon^{2 n-1}}\bigg\vert_\mu,
\end{equation}
where $a_n$ is related to zeta function by
\begin{equation}
a_n = 2 (1 - 2^{1-2 n})\,\zeta(2 n).
\end{equation}
A complete expansion in $\theta$ needs to include the expansion of the chemical potential $\mu$ too. To this end, the density of states (DOS) $n(\varepsilon)$ is introduced such that
\begin{equation}
\label{eq:DOS}
\dfrac{N_e}{V} = \int_{-\infty}^\infty n(\varepsilon) f(\varepsilon)\, d\varepsilon,
\end{equation}
where $N_e$ is the total number of electrons in a volume $V$. One can then show that, at second order in $\theta$,
\begin{equation}
\label{eq:mu}
\mu = \varepsilon_F \, \left(1 - \dfrac{\pi^2}{6} \dfrac{\varepsilon_F n'(\varepsilon_F)}{n(\varepsilon_F)}\, \theta^2\right),
\end{equation}
provided that the DOS $n(\varepsilon)$ does not vary from 0\,K to $T$.
As a consequence, $\mu$ can be replaced by $\varepsilon_F$ in every term of second order in $\theta$. In the same spirit, one gets
\begin{equation}
\int_{-\infty}^\mu g(\varepsilon) \, d\varepsilon = \int_{-\infty}^{\varepsilon_F} g(\varepsilon) \, d\varepsilon - \dfrac{\pi^2}{6} \dfrac{\varepsilon_F n'(\varepsilon_F)}{n(\varepsilon_F)}\,g(\varepsilon_F)\, \theta^2.
\end{equation}
We sum up with the expression of Sommerfeld expansion at second order in $\theta$
\begin{eqnarray}
\label{eq:Sommerfeld}
&\int_{-\infty}^\infty g(\varepsilon) f(\varepsilon)\, d\varepsilon ~ &= \int_{-\infty}^{\varepsilon_F} g(\varepsilon) \, d\varepsilon  \\ \nonumber
& &+ \dfrac{\pi^2}{6} \varepsilon_F^2 \left(g'(\varepsilon_F)- \dfrac{n'(\varepsilon_F)}{n(\varepsilon_F)}\,g(\varepsilon_F)\right)\, \theta^2.
\end{eqnarray}
As in the case of Eq.\,\eqref{eq:mu}, the function $g(\varepsilon)$ is assumed to stay unchanged from 0\,K to $T$. Otherwise, there is an implicit dependence on temperature in every term involving $g$ or its derivative. 

\subsection{Sommerfeld-like expansion of entropy}
\label{Sommerfeld_S}

A Sommerfeld-like expansion of the electronic entropy, as given by Eq.\,\eqref{eq:S}, can be performed using the function $s(\varepsilon)$ in lieu of the FD derivative $f'(\varepsilon)$.
Provided that the DOS $n(\varepsilon)$ is not too rapidly varying in the neighborhood of the chemical potential, it can be Taylor-expanded within the integral. At second order, one gets
\begin{equation}
n(\varepsilon) = n(\mu) + (\varepsilon - \mu) \, n'(\mu) + \frac{1}{2} (\varepsilon - \mu)^2 \, n''(\mu).
\end{equation}
Again, the term of order 1 does not contribute since the function $s(\varepsilon)$ is symmetric about the value of the chemical potential $\mu$
\begin{equation}
\dfrac{S}{V k_B} = - n(\mu)\,\int_{-\infty}^{+\infty} s(\varepsilon) \, d\varepsilon - \frac{1}{2} \,  n''(\mu)\,\int_{-\infty}^{+\infty}  (\varepsilon - \mu)^2 \,s(\varepsilon) \, d\varepsilon. 
\end{equation}
The integrals are evaluated using the same change of variable as before
\begin{equation}
z = \beta (\varepsilon - \mu),
\end{equation}
leading to
\begin{equation}
\dfrac{S}{V k_B} = b_1\, n(\mu)\,k_B T + b_3 \,   n''(\mu)\,(k_B T)^3, 
\end{equation}
with (see Appendix \ref{appD}):
\begin{equation}\label{b1}
b_1 = 2\,\int_{-\infty}^{+\infty} \log(1+e^z)/(1+e^z)\,dz = \dfrac{\pi^2}{3},
\end{equation}
and
\begin{equation}\label{b3}
b_3 = \int_{-\infty}^{+\infty} z^2\, \log(1+e^z)/(1+e^z)\,dz = \frac{7\pi^4}{90}.
\end{equation}
More generally, for the $(k_BT)^{2n+1}$ term in the expansion, one is faced with the evaluation of the integral
\begin{equation}
\int_{-\infty}^{+\infty} z^{2n}\, \log(1+e^z)/(1+e^z)\,dz.
\end{equation}
As before, to account for the temperature dependence of the chemical potential $\mu$, we use Eq.\,\eqref{eq:mu} to get a Sommerfeld-like expansion at third order

\begin{eqnarray}
\label{eq:S_S}
&\dfrac{S}{V k_B}\, &= b_1 \,\varepsilon_F\, n(\varepsilon_F) \, \theta \\
& &+ \varepsilon_F^3\,\left(\dfrac{\pi^2}{6}\,b_1\, \dfrac{\left[n'(\varepsilon_F)\right]^2}{n(\varepsilon_F)} +b_3\, n''(\varepsilon_F) \right) \, \theta^3.\nonumber
\end{eqnarray}

We shall investigate in the following results the domain of validity of this Sommerfeld expansion. One can conjecture that it breaks down when the DOS $n(\varepsilon)$ does not vary smoothly in the vicinity of the chemical potential value $\mu$. In particular, when the AA model predicts that the atom is neutral at sufficiently low density and temperature, there is only bound states in the DOS with Dirac distribution according to Eq.\,\eqref{eq:DOS_AA}. In App.\,\ref{NAG}, a discussion is provided about the AA predictions in this limit of a neutral atomic gas. In other circumstances, as the density increases at low temperature, pressure ionization can occur that leads to resonances in the DOS and one can wonder whether this impedes or not the validity of the Sommerfeld expansion.

\subsection{Free electron gas}
\label{FEG}

There is a solution to our problem which has met great success in solid-state physics: the FEG model. Sommerfeld proposed this model to improve Drude's model of conductivity \cite{Sommerfeld1928, Ashcroft1976}. He assumed that the interaction between the ions and the valence electrons can be neglected, the ions being only responsible for the charge neutrality in the metal. He further assumed that the interactions between electrons can be ignored as a result of screening effects. More importantly, he stressed that the Pauli exclusion principle requires that each quantum state of the system can only be occupied by a single electron, this restriction of available electron states being taken care by FD statistics. 

For our purpose, we shall only need the non-relativistic version of this model. Actually, the relativistic effects are only important for bound states at low temperature. The relativistic formulation is more involved and can be found in Faussurier's paper \cite{Faussurier2016a}. 

In the FEG model, the Fermi energy depends on the electron density $n_e$, which is uniform in the model, through

\begin{equation}
\varepsilon_F = \dfrac{1}{2}\,(3 \pi^2 \, n_e)^{2/3}.
\end{equation}
Defining a mean ionization $Z^*$ by $n_e = Z^*\,n_i = Z^*/V$, one can see that $\varepsilon_F \propto (Z^* \rho)^{2/3}$ in this model. The DOS $n(\varepsilon)$ reads 
\begin{equation}
n(\varepsilon) = \dfrac{3}{2}\,\dfrac{n_e}{\varepsilon_F}\,\sqrt{\dfrac{\varepsilon}{\varepsilon_F}} = \dfrac{1}{\pi^2}\sqrt{2\varepsilon},
\end{equation}
leading to the Sommerfeld expansion at first order in $\theta = k_B T / \varepsilon_F$
\begin{equation}
\label{eq:FEG}
S = \frac{3}{2}\,k_B \, b_1\,Z^*\,\theta.
\end{equation}
To sum up, the entropy per atom $S$ is proportional to $(Z^*)^{1/3}\,\rho^{-2/3}\,T$ at first order in the Sommerfeld expansion within the FEG model.

\section{Results}
\label{rst}
We have adapted to {\sc Inferno}-like modeling an AA  code developed by Jean Bruneau to study the model initially proposed by Rozsnyai \cite{Rozsnyai1972}, where the discrete bound states are broadened into canonical bands, and recently improved by Massacrier \cite{Massacrier2021} and Callow \cite{Callow2023}, where the free states are computed quantum-mechanically instead of treated within the semi-classical TF approximation used by Rozsnyai. Actually, Bruneau's code differs from these recent works by the use of the Dirac equation instead of the Schrodinger equation. We named {\sc Nirvana} this {\sc Inferno}-like adaptation of Bruneau's code.

Using {\sc Nirvana}, we computed the entropy of hydrogen (H, $Z = 1$, $A = 1.00784$,  belonging to the $s-$blocks of the periodic table), aluminum (Al, $Z = 13$, $A = 26.98$, belonging to the $p-$blocks), iron (Fe, $Z = 26$, $A = 55.845$, belonging to the $d-$blocks), and cerium (Ce, $Z = 58$, $A = 140.12$, belonging to the $f-$blocks).
The thermodynamic conditions of temperature $T$ and density $\rho$ cover the range from $0.01$ to $10$\,eV and $10^{-4}$ to $10^{4}$ g.cm$^{-3}$. The $T-\rho$ grid is logarithmic with 6 points per decade in $T$ and $\rho$, at: 1., 1.47, 2.15, 3.16, 4.64,  6.81. For each element, two figures show the two facets provided by isochores and isotherms. Since the entropy is an increasing function of temperature, the isotherms are in decreasing order of temperature from top to bottom. As for the isochores, they are in increasing order of density from top to bottom, since the entropy is a decreasing function of the density. We used Kohn-Sham XC-functional \cite{Kohn1965}. The sensitivity of the results to the choice of XC-functional is investigated in Sec.\,\ref{xc}.

\begin{figure}[!ht]
\begin{center}
\includegraphics[clip,width=1.\columnwidth]{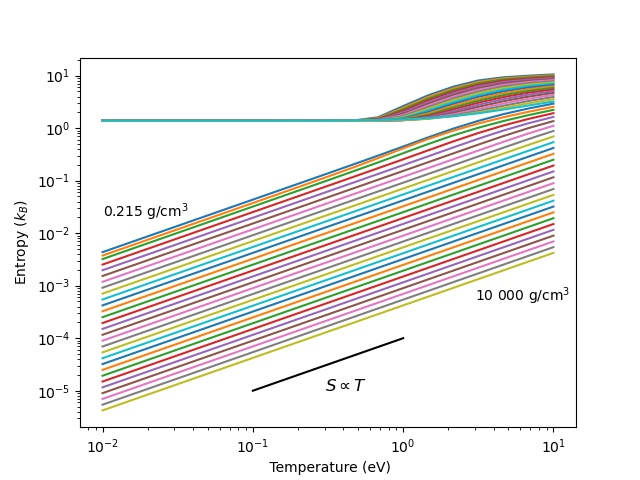}
\includegraphics[clip,width=1.\columnwidth]{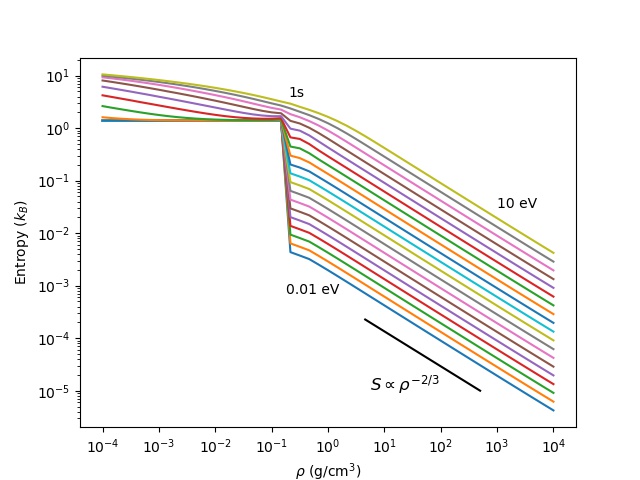}
\end{center}
\caption{Isochores (top) and isotherms (bottom) of electronic entropy per atom obtained for hydrogen using the {\sc Inferno}-like model implemented in the {\sc Nirvana} code. There are 6 points per decade in $T$ and $\rho$, at: 1., 1.47, 2.15, 3.16, 4.64,  6.81.  The range in density of pressure ionization of the $1s$ state   is indicated with its label in bottom panel.}
\label{fig:H}
\end{figure}

\subsection{Hydrogen}
\label{H}
%
With its simple electronic structure composed of only one electron per atom, hydrogen exhibits already the main features that are met for other elements. It is clear from Fig.\,\ref{fig:H} that there are two regimes: at low enough density, the AA model predicts that the atom is neutral with only bound states, whereas, at high density, there is a continuum of free states sitting on a fully occupied electronic configuration, just the nucleus in the hydrogen case. The transition between both regimes is rather abrupt, here at around $0.1-0.2$ g.cm$^{-3}$, with a WS radius of the confinement sphere $r_\text{ws} = 2-3\,a_0$, where $a_0$ is the Bohr radius. This is reminiscent of Mott's prediction of metal--insulator transition when the number of free electrons within a sphere of radius equal to $a_0$ vanishes \cite{Mott1949}. Before this transition, there is a slight increase of the entropy values with increasing density (see bottom panel of Fig.\,\ref{fig:H}). This nonphysical behavior is to be attributed to the possible lack of thermodynamic consistency discussed in Sec.\, \ref{INFERNO}.

The low-density regime can be described as composed of neutral atoms until the temperature reaches values higher than around 0.5\,eV. The entropy of the neutral atom is equal to 1.39\,$k_B$ per atom in accordance with the analysis presented in App.\,\ref{NAG} applied to hydrogen with only one electron in the $1s_{1/2}$ bound state. For temperature higher than 0.5\,eV, there is a thermal electronic contribution to the EOS, which increases with the ionization triggered by the rise of temperature. At these high temperatures, the entropy along isochores does not depend on temperature linearly, and the Sommerfeld expansion is not valid.

In the high-density regime, one observes the linear dependence of entropy with temperature predicted by the Sommerfeld expansion (see top panel of Fig.\,\ref{fig:H}). The coefficient of the temperature in the expansion depends on density as a power law around $\rho^{-2/3}$ (see bottom panel of Fig.\,\ref{fig:H}). These dependencies on temperature and density suggest that compressed hydrogen behaves as a FEG (see Sec.\,\ref{FEG}). We shall see in Sec.\,\ref{xc} that this is only approximate. Interestingly, the density power law does not apply to the intermediate density regime between $0.2$ and $0.4$ g.cm$^{-3}$, which presents a weaker slope due to the dissolution of the resonant feature, associated with the ionized $1s_{1/2}$ state, in the DOS of free states.

\begin{figure}[!ht]
\begin{center}
\includegraphics[clip,width=1.\columnwidth]{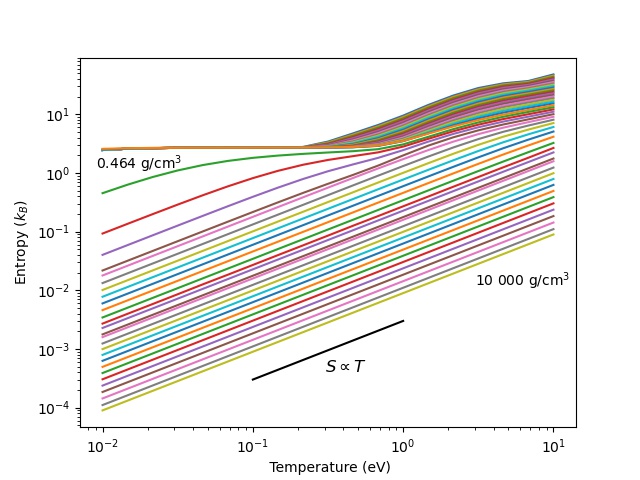}
\includegraphics[clip,width=1.\columnwidth]{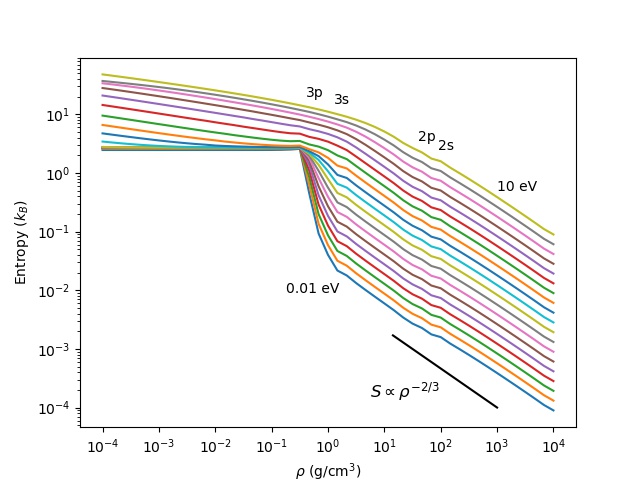}
\end{center}
\caption{Isochores (top) and isotherms (bottom) of electronic entropy per atom obtained for aluminum using the {\sc Inferno}-like model implemented in the {\sc Nirvana} code. There are 6 points per decade in $T$ and $\rho$, at: 1., 1.47, 2.15, 3.16, 4.64,  6.81. The range in density of pressure ionization of each bound state   is indicated with the state label in bottom panel.}
\label{fig:Al}
\end{figure}

\subsection{Aluminum}
\label{Al}

The results for aluminum are presented in Fig.\,\ref{fig:Al}. For one of the isochores at $\rho = 31.6$ g.cm$^{-3}$, characteristic of the pressure ionization of the $2p_{1/2}$ and $2p_{3/2}$ states, the {\sc Nirvana} code did not reach convergence. This engages six electrons which transit from the bound states to associated resonances from one iteration to the following, during the self-consistent-field search for convergence. The numerics is toughly stressed since the chemical potential $\mu$ varies strongly according to the configurations with bound states or resonances, and a strong requirement of discretization around $\mu$ is required. We did not pursue the numerical effort to resolve this issue since there exists ways to circumvent it by a broadening of the discrete bound states using canonical bands \cite{Rozsnyai1972, Massacrier2021, Callow2023} or more rigorously, albeit involved, using the Green function approach extending the eigenvalue problem to the complex plane, as introduced by Starrett \cite{Starrett2015, Starrett2019}. Instead, we interpolated the results of the nearest converged isochores at 27.5 and 33.5 g.cm$^{-3}$.
 
The transition from the low-density regime to the high-density one occurs between 0.3 and 0.7 g.cm$^{-3}$, corresponding to WS radii between 5 and 6\,$a_0$.
 
The low-density regime can be described as composed of neutral atoms with the following configuration: 
$[\text{Ne}]\,3s_{1/2}^2$\\$\,3p_{1/2}^{0.36}\,3p_{3/2}^{0.64}$. 
In this case, the level $3p_{1/2}$ and  
$3p_{3/2}$ are close in energy, $-0.0719$ and $-0.0713$\, a.\,u.  respectively, and their FD factor is almost the same close to $1/6$. Assuming a non-relativistic configuration with only one electron on the $3p$ shell,  the analysis presented in App.\,\ref{NAG} predicts a value of the entropy per atom of 2.7\,$k_B$ in accordance with the result of {\sc Nirvana}. For temperature higher than 0.2\,eV, there is a thermal electronic contribution to the EOS, which does not depend on temperature linearly as in the Sommerfeld expansion. 
 
In the high-density regime, the  entropy depends linearly on temperature as predicted by the Sommerfeld expansion (see top panel of Fig.\,\ref{fig:Al}). The function, that multiplies the temperature in this expansion, seems to depend on both density and ionization. In the bottom panel of Fig.\,\ref{fig:Al}, each isotherm exhibits a plateau in the same small density range whenever pressure ionization occurs. In these plateaus, the entropy is more or less constant as a result of the progressive dissolution of the associated resonance.
In between these plateaus, the entropy follows the power law around $\rho^{-2/3}$ of the FEG at a level that seems to adjust to the new ionization as suggested by the FEG model (see Eq.\,\eqref{eq:FEG} in Sec.\,\ref{FEG}).  Here, the AA model predicts shell effects that are not accounted for by the FEG model. Elucidating whether these shell effects are realistic or artifacts deserves further investigation.

\begin{figure}[!t]
\begin{center}
\includegraphics[clip,width=1.\columnwidth]{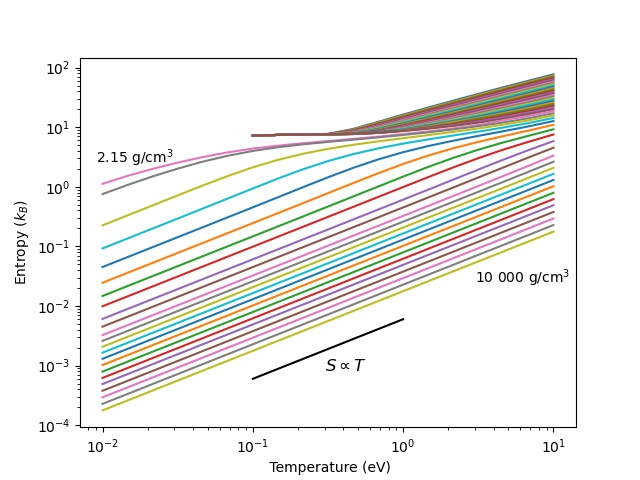}
\includegraphics[clip,width=1.\columnwidth]{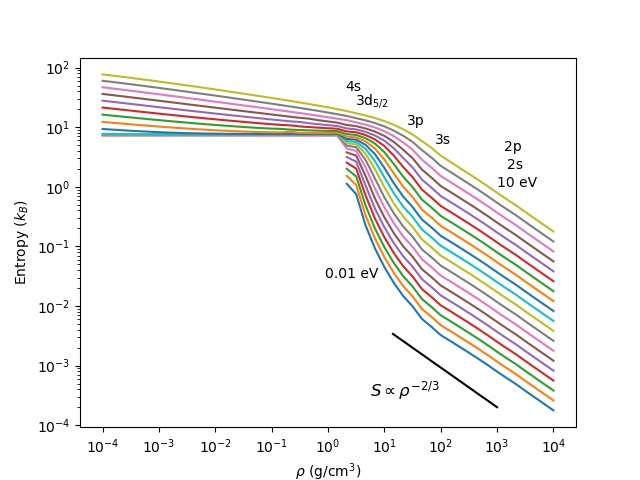}
\end{center}
\caption{Isochores (top) and isotherms (bottom) of electronic entropy per atom obtained for iron using the {\sc Inferno}-like model implemented in the {\sc Nirvana} code.  There are 6 points per decade in $T$ and $\rho$, at: 1., 1.47, 2.15, 3.16, 4.64,  6.81.  The range in density of pressure ionization of each bound state   is indicated with the state label in bottom panel.}
\label{fig:Fe}
\end{figure}

\begin{figure}[t!]
\begin{center}
\includegraphics[clip,width=1.\columnwidth]{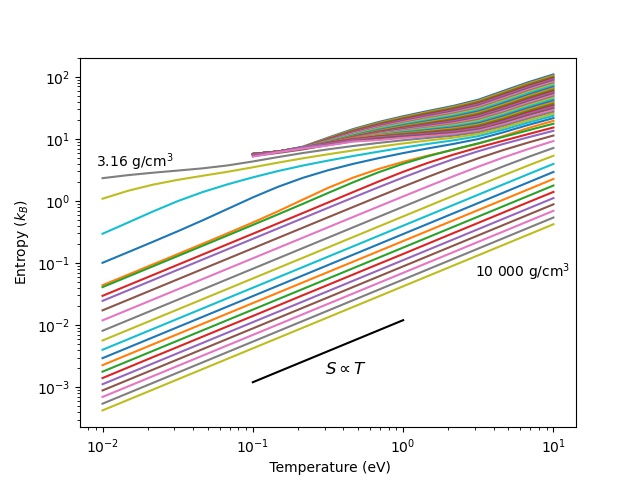}
\includegraphics[clip,width=1.\columnwidth]{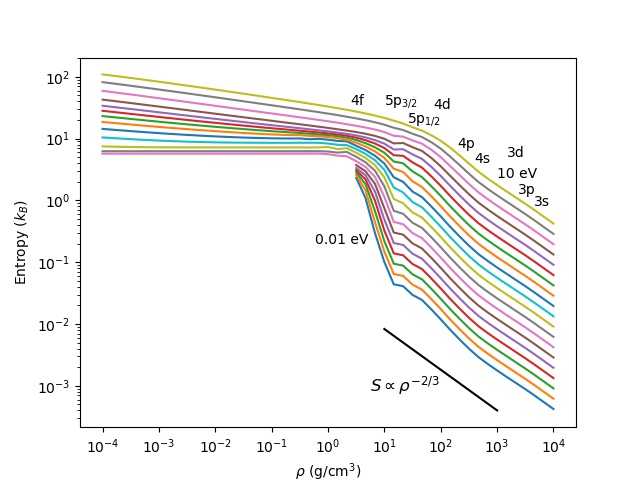}
\end{center}
\caption{Isochores (top) and isotherms (bottom) of electronic entropy per atom obtained for cerium using the {\sc Inferno}-like model implemented in the {\sc Nirvana} code.  There are 6 points per decade in $T$ and $\rho$, at: 1., 1.47, 2.15, 3.16, 4.64,  6.81.  The range in density of pressure ionization of each bound state   is indicated with the state label in bottom panel.}
\label{fig:Ce}
\end{figure}

\subsection{Iron}
\label{Fe}

The results for iron are presented in Fig.\,\ref{fig:Fe}. In this case, the AA model, as implemented in the {\sc Nirvana} code, is faced with an issue related to the convergence of the neutral atom configuration at low temperature. At a temperature of 0.1\,eV, the converged configuration is 
$[\text{Ar}]\, 3d_{3/2}^{3.15}\, 3d_{5/2}^{3.54}\,4s_{1/2}^{1.31} $  
with a value of the entropy of $7.41\,k_B$. The three last partially occupied levels are located at close energies of $-0.1452$, $-0.1400$, and $-0.1415$\,a.\,u. A fictitious "degenerate" level, assembling these three levels, of degeneracy 12 with occupancy 7, leads to a value of the entropy of $8.15\,k_B$ according to the analysis of App.\,\ref{NAG}. At temperatures lower than 0.1\,eV, the spread around the chemical potential in the FD distribution lessens and the three levels are much more into competition. Each preferred level leads to a different screening of the nucleus charge favoring another level, and so on. As a matter of fact, as discussed in App.\,\ref{NAG}, the best method to determine the fundamental state of an isolated atom requires a multi-configuration-interaction calculation, that cannot compete with the DFT formulation. So, we left aside the corresponding parts of the isochores, namely for densities less than 1.47 g.cm$^{-3}$ and temperatures less than 0.1\,eV.

The transition from the low-density regime to the high-density one occurs between 1.5 and 2.3 g.cm$^{-3}$, corresponding to WS radii between 4.0 and 4.6\,$a_0$.
 
As for the other elements, ionization shows up in the low-density regime above a temperature of about 0.3\,eV. It produces a thermal electronic contribution to the EOS, which again does not depend on temperature linearly as in the Sommerfeld expansion.

In the high-density regime, the entropy depends linearly on temperature as predicted by the Sommerfeld expansion (see top panel of Fig.\,\ref{fig:Fe}). In the bottom panel of Fig.\,\ref{fig:Fe}, the values of the entropy decrease with increasing density with a higher slope than the power law of $\rho^{-2/3}$ characteristic of the FEG, along each isotherm. It is only for density higher than 100 g.cm$^{-3}$ that the FEG behavior is recovered.

\subsection{Cerium}
\label{Ce}

The results for cerium are presented in Fig.\,\ref{fig:Ce}. In this case, the AA model, as implemented in the {\sc Nirvana} code, is faced with both issues met with aluminum and iron. As with iron, the convergence of the neutral atom configuration at low temperature is only reached at a temperature of 0.1\,eV. Therefore, for the same reasons, we left aside the data for densities less than 3.16 g.cm$^{-3}$ and 0.1\,eV. As with aluminum, one of the isochores at $\rho$ = 1470 g.cm$^{-3}$ presents convergence issues related here to the pressure ionization of the $3d_{3/2}$ bound state. So, we interpolated the results of the nearest isochores at 1400 and 1600 g.cm$^{-3}$.

The transition from the low-density regime to the high-density one occurs between 2 and 3 g.cm$^{-3}$, corresponding to WS radii between 5. and 6.\,$a_0$.
 
The DFT approach used in the AA model predicts a value of the entropy for the neutral atom at 0.1\,eV of 5.71\,$k_B$ per atom. At a temperature of 0.1\,eV, the converged configuration is 
$[\text{Xe}]\, 4f_{5/2}^{1.36}\, 4f_{7/2}^{0.11}\, 5d_{3/2}^{0.41}\, 5d_{5/2}^{0.12}\,6s_{1/2}^{2}$. Here, there are four partially occupied levels with close energies of $-0.0709$, $-0.0597$, $-0.0675$, and $-0.0613$\, a.\,u., respectively. Assembling these four levels into a fictitious level of degeneracy 24 with occupancy 2, leads to a value of the entropy of $6.88\,k_B$ according to the analysis of App.\,\ref{NAG}. Ionization shows up in this low-density regime above a temperature of about 0.2\,eV, without a linear dependence in temperature as in Sommerfeld model.

In the high-density regime, the entropy depends linearly on temperature as predicted by the Sommerfeld expansion (see top panel of Fig.\,\ref{fig:Ce}). In the bottom panel of Fig.\,\ref{fig:Ce}, the values of the entropy decrease with increasing density with a higher slope than the power law around $\rho^{-2/3}$ of the FEG, along each isotherm. It is only for density higher than 250 g.cm$^{-3}$ that the FEG behavior is  recovered.

\begin{figure*}[t!]
\begin{center}
\includegraphics[clip,width=0.7\textwidth]{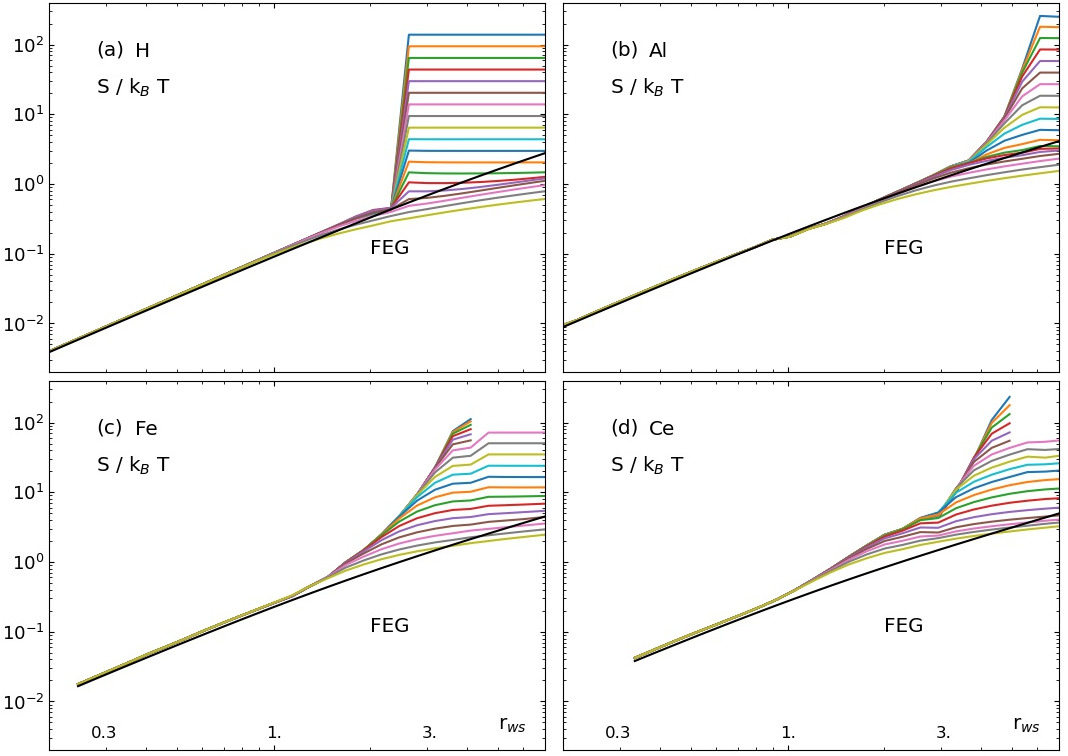}
\end{center}
\caption{Isotherms of electronic entropy $S$ in $k_B$ per atom divided by $T$ in eV as a function of $r_\text{ws}$ obtained for hydrogen (a), aluminum (b), iron (c), and cerium (d), using the {\sc Inferno}-like model implemented in the {\sc Nirvana} code. Comparison is also presented with the FEG using More's fit of the Thomas-Fermi ionization (black lines).}
\label{fig:Tdiv}
\end{figure*}

\section{Analysis}
\label{model}

\subsection{Sommerfeld approximation}
\label{Sommerfeld_T}

As a global exploitation of the results presented in Sec.\,\ref{rst}, we check the linear dependence on temperature of the electronic entropy in Fig.\,\ref{fig:Tdiv}, where each isotherm is divided by the corresponding temperature, for each element. The curves are plotted as functions of the WS radius $r_\text{ws}$ to get rid of the unnecessary dependence on the molar mass $A$. Then, the entropy is an increasing function of the WS radius. In the considered conditions, the lowest isotherm corresponds to the highest temperature of 10\,eV when the entropy is divided by temperature. Respectively, the highest isotherm corresponds to the lowest temperature of 0.01\,eV.  To compare easily the levels and slopes of the isotherms, we keep the same frame for all panels, with the same ranges of WS radii and electronic entropy values. We added the prediction of the FEG model using the Thomas-Fermi (TF) ionization, given by More's fit \cite{More1985}. We checked that the ionization given by {\sc Nirvana} does not change these predictions. This is due to the weak dependence on $Z^*$ of the FEG model.

The linear dependence on temperature of Sommerfeld's model is observed at low temperatures in the high-density regime, for values of the WS radii $r_\text{ws}$ lower than 2 to 6 Bohr radius ($a_0$) according to where each element transits to a neutral system. The FEG model with TF ionization does a good job up to values of $r_\text{ws}$ around 1, in a highly compressed regime. For higher values of $r_\text{ws}$, lower densities, {\sc Nirvana} results depart from the FEG predictions more or less depending on the element. In this range of $r_\text{ws}$ values, above $r_\text{ws} = 1$, the computed isotherms leave the Sommerfeld's prediction progressively, the highest temperatures at a lower $r_\text{ws}$ value than the lowest temperatures. As a metric for the temperature domain of validity of the Sommerfeld expansion, we propose to use the degeneracy ratio $\theta = k_B T / \varepsilon_F$ using the TF ionization for the evaluation of the Fermi energy $\varepsilon_F$, as in the FEG model. A visual inspection of Fig.\,\ref{fig:Tdiv} allows one to  keep track of the values of $r_\text{ws}$ where each isotherm moves away from the common line by 10\% in entropy value. This indicates that the linearity of the electronic entropy with temperature ceases above $\theta = 0.4$ for hydrogen, 0.3 for aluminum, and  0.05 for iron and cerium. Based on these results, we propose to adopt a conservative limit of $\theta = 0.05$ for all elements.  Needless to say that this prescription requires further investigation. Together with the limit in WS radius $r_\text{ws}$ of the metal-insulator transition, the limit in degeneracy ratio $\theta$ allows one to study the Sommerfeld expansion of the electronic entropy for a given chemical element from only one isotherm at sufficiently low temperature. We pursue this line of investigation in Sec.\,\ref{Z} where trends with the atomic number $Z$ are studied.

\begin{figure}[t!]
\begin{center}
\includegraphics[clip,width=1.\columnwidth]{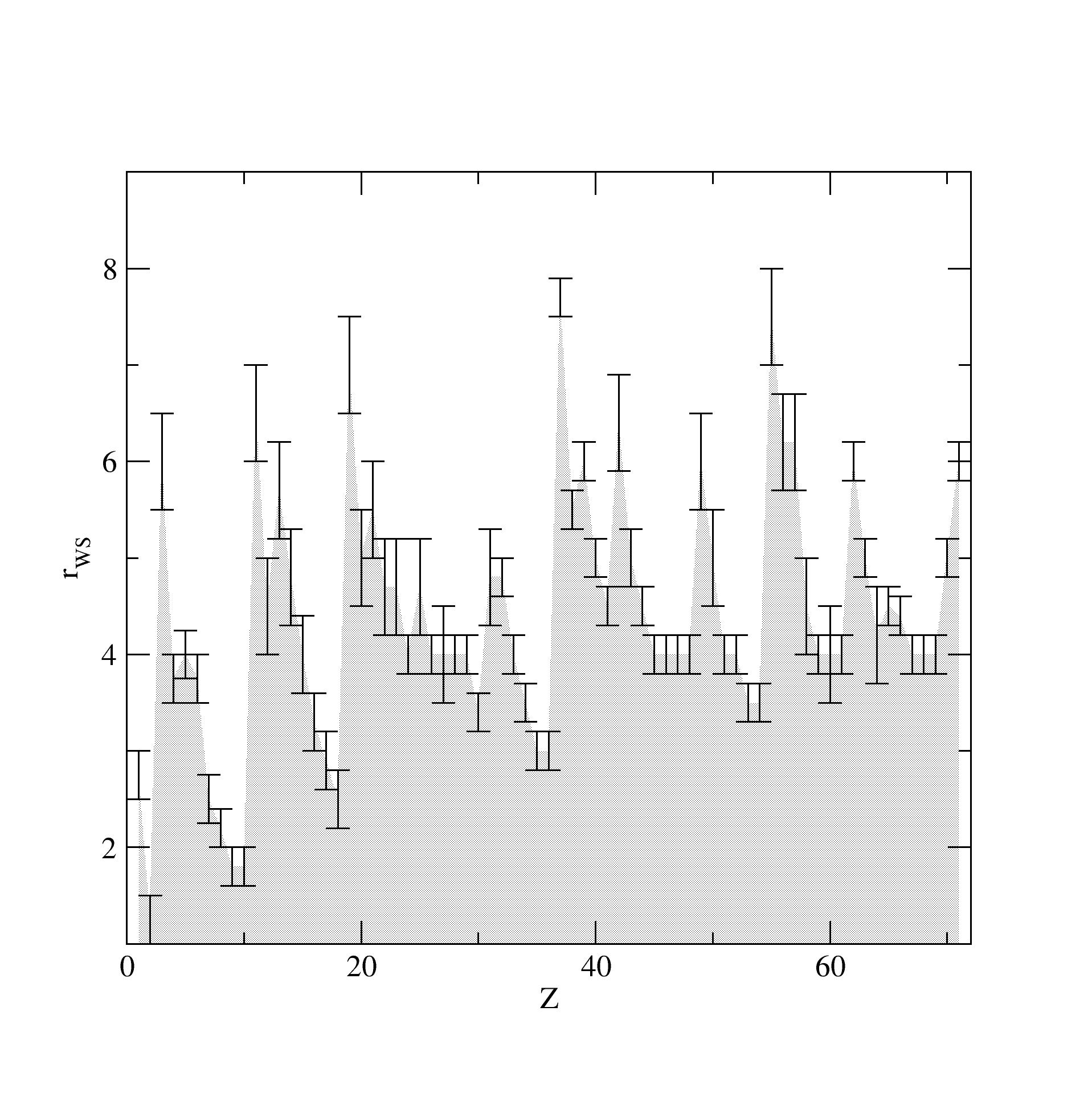}
\end{center}
\caption{WS radii at metal-insulator transition in {\sc Nirvana} as a function of the atomic number $Z$.}
\label{fig:rws_limit}
\end{figure}

\subsection{Z trends}
\label{Z}
The preceding analysis of Sec.\,\ref{Sommerfeld_T} suggests to determine the density-dependent coefficient of  temperature in the Sommerfeld expansion for all chemical elements using just one isotherm at sufficiently low temperature. To set up this temperature, we decided to look for a value $T$ which  leads to degeneracy parameter $\theta$ less than the limit of 0.05 in the whole range of WS radius values $r_\text{ws}$ of the high-density regime, which is limited from above by the metal-insulator transition. Since  $\varepsilon_F \propto (Z^* )^{2/3} / r_\text{ws}^2$, the highest value of $\theta$ is obtained from the lowest value of $\varepsilon_F$, using $Z^* = 1$ and the highest value of $r_\text{ws}$ for the metal-insulator transition. We shall see that this maximum  $r_\text{ws}$ value is close to 8. (see Fig.\,\ref{fig:rws_limit}) leading to a value of the Fermi energy of $\varepsilon_F$ around 0.8\,eV. The limit temperature corresponding to $\theta = 0.05$ is then $T = 0.04$\,eV. We adopted $T= 0.01$\,eV as a conservative choice. The corresponding isotherm was computed in the density range from 0.1 to 10$^4$ g.cm$^{-3}$ for the elements of atomic number $Z$ in the range from 1 (hydrogen) to 71 (lutetium).

\subsubsection{Metal -- insulator transitions}
\label{Mott}

We determined the value of the WS radius $r_\text{ws}$ at the metal-insulator transition by visual inspection for each element. As Fig.\,\ref{fig:rws_limit} shows, there is no trend following the atomic number $Z$, except for the natural sequence following the periodic table of the elements. Overall, a typical limit of $r_\text{ws} = 4 \,a_0$ obtrudes. As we shall see in Sec.\,\ref{xc}, these values of the critical WS radius are highly sensitive to the XC functional used.

\begin{figure}[t!]
\begin{center}
\includegraphics[clip,width=1.\columnwidth]{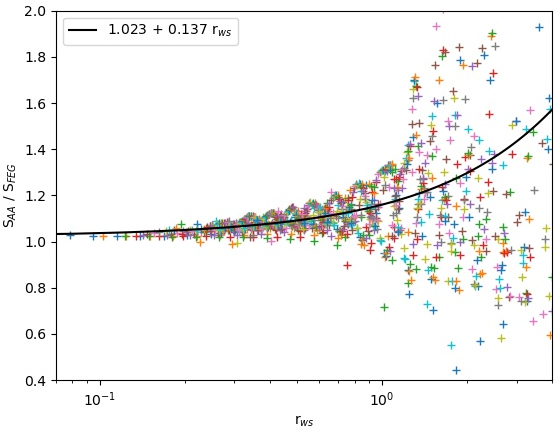}
\end{center}
\caption{Ratios between the electronic entropy $S$ predicted by {\sc Nirvana} and the FEG model at 0.01\,eV for elements from $Z$ = 1 (hydrogen) to 71 (lutetium) as a function of the WS radius. Also plotted is a linear regression in the range of $r_\text{ws}$ values from 0 to 1.}
\label{fig:Z_FEG_bis}
\end{figure}

\subsubsection{FEG regimes}
\label{powerlaw}

We checked the validity domain of the FEG model by computing for each element the ratio between the electronic entropy values computed with {\sc Nirvana} and predicted by the FEG model as a function of the WS radius, for the isotherm at 0.01\,eV. The results are gathered in Fig.\,\ref{fig:Z_FEG_bis}. For WS radii $r_\text{ws}$ less than 1, the spread among the results is less than 20\% with a tendency for higher electronic entropy than predicted by the FEG model as the value of 1 is reached. An empirical correction of linear form, $c_0 + c_1 r_\text{ws}$, with $c_0 = 1.023$ and $c_1 = 0.137$, can multiply the FEG formula for the entropy $S$ leading to an accuracy of $\pm\,10$\% in this range. For WS radii $r_\text{ws}$ greater than 1, we did not observe any trends with the atomic number $Z$. In this range, accurate results can only be obtained from the AA code. We provide a small database in supplemental materials to satisfy this need. Finally, although the FEG model does not completely reproduce the AA results at all WS radii, it flattens out the main trend of increase in electronic entropy $S$ as the density $\rho$ decreases, or alternatively the WS radius $r_\text{ws}$ increases. Comparing Figs.\,\ref{fig:Tdiv} and \ref{fig:Z_FEG_bis}, the variations of $S$ no longer spread over decades. We shall use this normalization of $S$ by the FEG predictions in Sec.\,\ref{xc} to highlight the sensitivity to XC functionals.

\begin{figure*}[t!]
\begin{center}
\includegraphics[clip,width=0.7\textwidth]{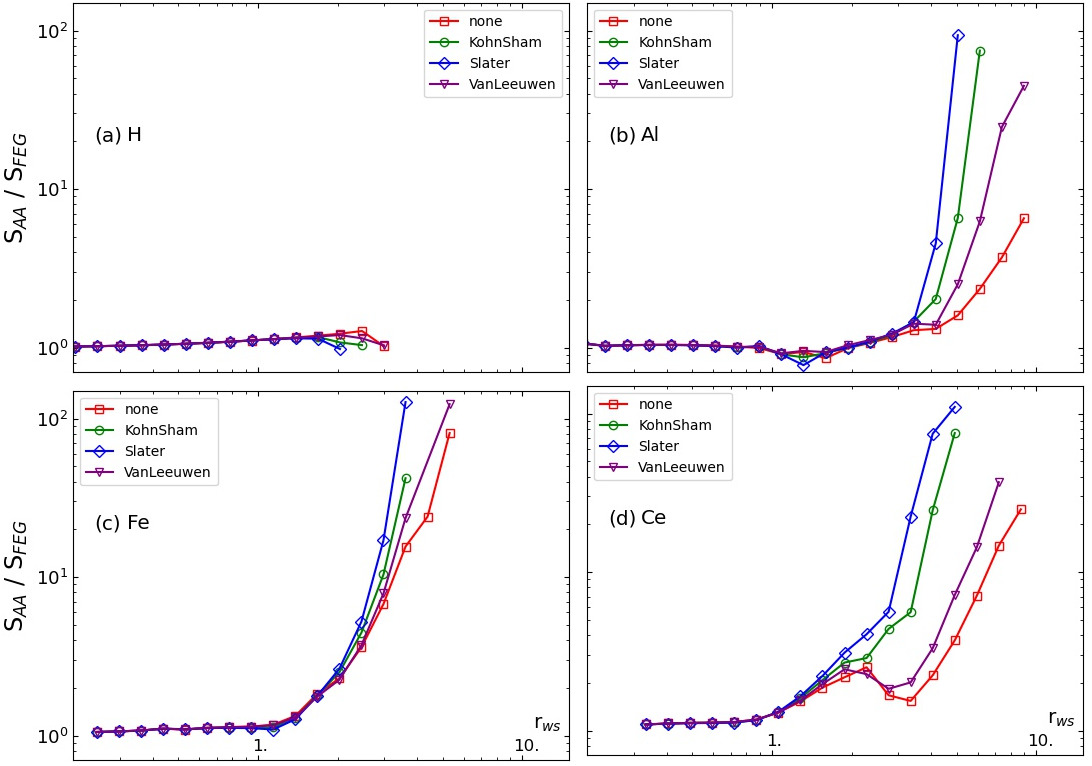}
\end{center}
\caption{Sensitivity to the XC functional of the ratio between the electronic entropy $S_{AA}$ predicted by {\sc Nirvana} and $S_{FEG}$  predicted by the FEG model at 0.01\,eV as a function of the WS radius $r_\text{ws}$. Four choices are compared: no XC functional ("none"), the Kohn-Sham functional ("KohnSham" \cite{Kohn1965}), the Slater functional ("Slater" \cite{Slater1951}), and the Van Leeuwen functional ("VanLeeuwen" \cite{Leeuwen1994}). Hydrogen is depicted in panel (a), aluminum in panel (b), iron in panel (c), and cerium in panel (d). Each curve is stopped at the WS radius where a metal-insulator transition occurs.}
\label{fig:Z_xc}
\end{figure*}

\subsection{sensitivity to xc}
\label{xc}

Here, we investigate the sensitivity of the AA predictions of electronic entropy $S$ to the choice of XC functionals used. We compare the choices of the Kohn-Sham functional used in the main text \cite{Kohn1965}, of the Slater functional \cite{Slater1951},  of the Van Leeuwen functional \cite{Leeuwen1994}, and the choice not to include any XC contribution. Although this set of functionals is limited, it is sufficient to highlight the main sensitivities. The used XC functionals are exchange-only functionals, neglecting the contribution from correlation. The Kohn-Sham and the Slater functionals belong to the family of X$\alpha$ functionals, with the following exchange potential
\begin{equation}
    V_{X\alpha}(r) = -\, \frac{3}{2} \,X_{\alpha} \left( \frac{3}{\pi} \right)^{{1}/{3}} \rho (r) ^{{1}/{3}},
\end{equation}
with $X_\alpha = 1$ for the Slater functional and $X_\alpha = 2/3$ for the Kohn-Sham functional. Both functionals use a local density approximation, whereas the Van Leeuwen functional includes corrections from the gradient of density, with a potential given by
\begin{equation}
    V_\text{VL}(r) =  -\, \beta \, \rho (r) ^{{1}/{3}} \, \dfrac{x^2}{1 + 3 \, \beta \, x \, \sinh^{-1}(x)},
\end{equation}
with the dimensionless quantity
\begin{equation}
    x = \dfrac{|\nabla \rho|}{\rho^{4/3}}.
\end{equation}
At contrast to other non-local approximations \cite{Jones2015, Yu2016}, the Van Leeuwen potential was built as a first attempt to preserve the asymptotic  Coulombic  behavior of the exchange potential, leading to a self-interaction correction.

The analysis of the preceding Secs.\,\ref{model}, \ref{Z}, and \ref{powerlaw}, prompts us to restrict the comparison between the results using different XC functionals to just one isotherm at 0.01\,eV. It also demonstrates that the ratio between the values of electronic entropy $S$ predicted by the AA model and the FEG one as a function of the WS radius $r_\text{ws}$ is most appropriated to highlight any discrepancies. In Fig.\,\ref{fig:Z_xc}, this ratio  is plotted as a function of the WS radius $r_\text{ws}$ for the preceding choices of XC functionals, at a constant temperature of 0.01\,eV, for hydrogen in the panel (a), aluminum in the panel (b), iron in the panel (c), and cerium in the panel (d). As expected for a weakly correlated electron system, the sensitivity to the XC choice is weak at high density, for WS radii less than 1 -- 2. For higher WS radii, at moderate density, discrepancies among the different XC choices show up. They are responsible of different values of the WS radius  $r_\text{ws}$ for the metal-insulator transition and different transitory behaviors in between  $r_\text{ws} \sim 2$ and this critical WS radius.

\section{Conclusion}
\label{conclusion}

The main purpose of this work was to contribute to the modeling of the electron-thermal part of the equation of state  in the low temperature regime, characteristic of solids and liquid metals. Among the different approaches used in the models for the multi-phase equation of state, the average atom is considered as the method providing the best compromise between accuracy, robustness, and computation time to obtain the electron-thermal contribution. Our analysis through the prism of Sommerfeld expansion lays a simplification, which can afford to avoid the pitfalls related to pressure ionization and the often associated discontinuities in thermodynamic variables. It also provides orders of magnitude of the sensitivity of the electronic entropy to modeling options.

Using an average-atom framework alike to the {\sc Inferno} model, we probed the validity of the Sommerfeld approximation to the electronic entropy at low temperature. We performed a thorough analysis of four different elements belonging to the $s-$, $p-$, $d-$, and $f-$ blocks of the periodic table, namely hydrogen, aluminium, iron, and cerium. This allowed us to delineate the validity domain of Sommerfeld approximation in temperature and density. The dimensionless ratio $\theta = k_B T / \varepsilon_F$ reducing the temperature to the Fermi energy  must be less than around 0.05 to preserve a linear dependence of the entropy on temperature. In this range of temperature, the coefficient of  temperature is a function of density, which can be investigated using just one isotherm at sufficiently low temperature. This allowed us to extend the analysis to a broader set of atomic numbers $Z$. The only trend with $Z$ is that the free electron gas model is within 10\% of the average-atom prediction at very high density, for Wigner-Seitz radius $r_\text{ws}$ less than 1, provided that a linear scaling in $r_\text{ws}$ is adopted with a 14\% increase at $r_\text{ws} = 1$.  

 At contrast with the semi-classical Thomas-Fermi average atom model, the quantum-mechanical version includes shell effects, but most importantly predicts that metal-insulator transition occurs although it is often missing from the multi-phase equation of state. The study of various elements with different atomic number $Z$ evidenced that a metal-insulator transition is predicted by the quantum-statistical average atom model. It occurs at different $r_\text{ws}$  values according to the element without any other trend than the filling of the periodic table. This transition is also very sensitive to the exchange and correlation functional used. Actually, the study of this transition is an active research field \cite{Calderin2011, Clerouin2012a, Korobenko2012a, Li2014, Minakov2018, Celliers2018a, Liu2019, Liu2020} that we only touched upon, since it is beyond  the scope of the present paper.

 As concern the best modeling choice for the electron-thermal part of the multi-phase equation of state, we propose to start from the Sommerfeld expansion of the electronic entropy as a foundation. The Sommerfeld expansion of the electronic entropy involves a density-dependent coefficient of the temperature.  Depending on the demand of accuracy for multi-phase equation of state, this coefficient is only approximately given by a power law in density. It is preferable to use the small database of the supplemental materials. We also showed that the free electron gas model is of limited value, restricted to very high densities.

We plan to enrich our study with \textit{ab initio} simulations to compare the electronic entropy values and the underlying density of states with the average atom ones. Preliminary results suggest that, although there is good agreement in the fluid phase the crystalline structure induces noticeable discrepancies, which should hopefully be taken into account by a rescaling of the average atom results from a limited number of simulations.


\section{References}
%
 \appendix
 
\section{Entropy route to thermal EOS}
\label{EOS}

Gibbs provided the following relation from the first principle of thermodynamics
\begin{equation}
dU = T dS -P dV + \mu dN.
\end{equation}
It expresses the energy conservation in systems in thermodynamic equilibrium where one can define the temperature $T$, the pressure $P$, and the chemical potential $\mu$. The variation of internal energy $U$ is connected to the variations of entropy $S$, volume $V$, and number of particles $N$. In the canonical ensemble, it is the free energy $A(T, V, N)$ that represents the thermodynamic potential through the following Legendre transform
\begin{equation}
A = U - T S,
\end{equation} 
leading to
\begin{equation}
dA = -S dT - P dV +\mu dN.
\end{equation}
At constant volume $V_0$ and number of particles $N_0$, the thermal free energy $\Delta A(T, V_0, N_0) = A(T, V_0, N_0) - A(0, V_0, N_0)$ reads
\begin{equation}
\Delta A(T, V_0, N_0) = - \int_0^T S(\tau, V_0, N_0) \, d\tau,
\end{equation}
and at the same time
\begin{equation}
\Delta A(T, V_0, N_0) = \Delta U(T, V_0, N_0) - T S(T, V_0, N_0),
\end{equation}
since $S(0, V_0, N_0)=0$ according to the third law of thermodynamics (Nernst theorem). The thermal internal energy $\Delta U(T, V_0, N_0) = U(T, V_0, N_0) - U(0, V_0, N_0)$ is then given as a functional of the entropy by
\begin{equation}
\Delta U(T, V_0, N_0) =  - \int_0^T S(\tau, V_0, N_0) \, d\tau + T S(T, V_0, N_0).
\end{equation}
The thermal pressure $\Delta P(T, V_0, N_0) = P(T, V_0, N_0) - P(0, V_0, N_0)$ involves a differentiation of the free energy, which translates into
\begin{equation}
\Delta P(T, V_0, N_0) = \int_0^T \dfrac{\partial S(\tau, V_0, N_0)}{\partial V_0}\bigg\vert_{\tau, N_0} \, d\tau,
\end{equation}
%

\section{Neutral atom gas}
\label{NAG}

The low-density and low-temperature regime can be described as composed of neutral atoms. In these conditions, there is no reason to add an electronic contribution to the EOS.
 Here, we just comment on the fact that the entropy value obtained within the AA framework for the neutral atom does not vanish as expected. 
 
Actually, to find the fundamental state of an isolated atom requires a multi-configuration computation where all degeneracy are reduced using configuration interaction to get the ultimate term representing the lowest possible state in energy of the atom. This calculation leaves no room to fluctuations and the value of the entropy should vanish. This is also in line with the absence of any electronic contribution to the EOS.
The finite temperature version of DFT cannot compete with such a kind of computation and only predicts an average configuration with possibly partially occupied shells that may lead to non-vanishing values of the entropy.

Within the AA model, when the density and the temperature are so low as to describe a system of neutral atoms, without free states, Eq.\,\eqref{eq:S} reduces to
\begin{equation}
\label{eq:S_H}
S = - k_B \sum_{s \text{\,bound}}  \left[f_s \,\log f_s + (1-f_s)\, \log(1- f_s)\right]\, X_s,
\end{equation}
where
\begin{equation}
X_s = 2 |\kappa_s| = 2 j_s + 1,
\end{equation}
since, at sufficiently low density, the WS radius is so large that the bound wave functions are completely included in the confinement volume and normalized within it
\begin{equation}
\int_0^{r_\text{ws}} \left[F_s^2(r) +G_s^2(r)\right] \, dr = 1.
\end{equation}
The remaining unknown is the chemical potential $\mu$ since
\begin{equation}
f_s = \dfrac{1}{1 + e^{\beta (\varepsilon_s -\mu)}}.
\end{equation}
It is given by Eq.\,\eqref{eq:AA_Z} which reads without free states as
\begin{equation}
\label{eq:Z_H}
Z = \sum_{s \text{\,bound}} f_s\, X_s =  \sum_{s \text{\,bound}} \dfrac{2 j_s + 1}{1 + e^{\beta (\varepsilon_s -\mu)}}.
\end{equation}
We also know that, at vanishing temperature, all the levels are fully occupied, from the lowest strongly-bound one up to the highest weakly-bound one, until the number of available electrons is exhausted. The Fermi energy $\varepsilon_F$ is then the energy of the last occupied state. Starting from this picture, we can figure out how the levels are occupied at low temperature. All the lowest levels, that can be fulfilled within the limit of $Z$ electrons, are occupied. The remaining $n_0$ electrons partially occupy the next level of degeneracy $X_0 = g_0$. Then, Eq.\,\eqref{eq:S_H} and  Eq.\,\eqref{eq:Z_H} reduce to
\begin{equation}
S = - k_B \left[f_0 \,\log f_0 + (1-f_0)\, \log(1- f_0)\right]\, g_0,
\end{equation}
and 
\begin{equation}
n_0 = f_0\,g_0 = \dfrac{g_0}{1 + e^{\beta (\varepsilon_0 -\mu)}}.
\end{equation}
The value of the chemical potential $\mu$ should be easily obtained unless there are more than one level candidate for the level $\varepsilon_0$ in the course of the iterations of the self-consistent field determination. We shall see that this kind of fluctuations occurs in some case.

There is a situation where the preceding reasoning cannot determine a value for the chemical potential $\mu$. It is when all the levels are fully occupied and there is no more electrons to distribute. In this case, the entropy $S$ vanishes and the chemical potential $\mu$ is undetermined.



\section{Proof of the values of coefficients $b_1$ and $b_3$}\label{appD}

We provide below a proof of the expressions of coefficients $b_1$ and $b_3$ defined respectively in Eqs. (\ref{b1}) and (\ref{b3}).
\subsection{Coefficient $b_1$}

We have to calculate the integral
\begin{equation}
b_1=\int_{-\infty}^{+\infty} \log(1+e^z)/(1+e^z)\,dz.
\end{equation}
Let us start with the change of variables $u=e^z$, yielding
\begin{equation}
b_1=\int_0^{\infty}\frac{\ln(1+u)}{u(1+u)}\,du.
\end{equation}
Let us now make the change of variable: $u=t/(1-t)$. One has $t=u/(1+u)$ and
\begin{equation}
b_1=-\int_0^{1}\frac{\ln(1-t)}{t}dt=\mathrm{Li}_2(1),  
\end{equation}
where
\begin{equation}
\mathrm{Li}_{s}(z)=\sum _{k=1}^{\infty }{z^{k} \over k^{s}}
\end{equation}
is the usual polylogarithm (${Li}_2$ is usually referred to as the ``dilogarithm''). One has in particular
\begin{equation}
\mathrm{Li}_{n}(1)=\zeta(n),
\end{equation}
where $\zeta$ is the Riemann zeta function. Therefore
\begin{equation}
b_1=\zeta(2)=\frac{\pi^2}{6}.
\end{equation}

\subsection{Coefficient $b_3$}

The determination of $b_3$ is more cumbersome. We have to calculate the integral
\begin{equation}
b_3=\int_{-\infty}^{+\infty} z^2\, \log(1+e^z)/(1+e^z)\,dz.
\end{equation}
Making the same successive two changes of variables $u=e^z$ and $t=u/(1+u)$ as for $b_1$, one gets
\begin{equation}
b_3=-\int_0^{1}\frac{\left[\ln (t)-\ln(1-t)\right]^2\ln(1-t)}{t}\,dt,  
\end{equation}
\emph{i.e.}
\begin{eqnarray}\label{b3bis}
&b_3=&-\int_0^{1}\frac{\ln^2(t)\ln(1-t)}{t}\,dt \\& &-\int_0^{1}\frac{\ln^3(1-t)}{t}\,dt +2\int_0^{1}\frac{\ln (t)\ln^2(1-t)}{t}\,dt. \nonumber
\end{eqnarray}
Lewin \cite{Lewin1981} provides the following expression (Eqs. (7.48) p. 199 and (7.61) p. 202):
\begin{eqnarray}\label{Lewin761}
&\mathrm{Li}_4(x)=\ln(x) \mathrm{Li}_3(x)&-\frac{1}{2}\ln^2(x)\mathrm{Li}_2(x)\\& &-\frac{1}{2}\int_0^x\ln^2(t)\frac{\ln(1-t)}{t}\,dt.\nonumber
\end{eqnarray}
From Eq. (\ref{Lewin761}) one obtains, taking the limit $x\rightarrow 1$:
\begin{equation}\label{first}
\int_0^1\frac{\ln^2(t)\ln(1-t)}{1-t}\,dt=-2\zeta(4).
\end{equation}
Using Eq. (7.62) p. 203 of Ref. \cite{Lewin1981} or integrating further by parts in Eq (\ref{Lewin761}) leads to:
\begin{eqnarray}\label{Lewin762}
&\mathrm{Li}_4(x)=&\ln(x) \mathrm{Li}_3(x)-\frac{1}{2}\ln^2(x)\mathrm{Li}_2(x)\\& &-\frac{1}{6}\ln^3(x)\ln(1-x)-\frac{1}{6}\int_0^x\frac{\ln^3(t)}{1-t}\,dt.\nonumber
\end{eqnarray}
Taking the limit $x\rightarrow 1$ in the latter equation and using the property that
\begin{equation}
\int_0^1\frac{\ln^3(t)}{1-t}\,dt=\int_0^1\frac{\ln^3(1-t)}{t}\,dt
\end{equation}
gives
\begin{equation}\label{second}
\int_0^x\frac{\ln^3(1-t)}{t}\,dt=-6\zeta(4).
\end{equation}
The last integral is the right-hand side of Eq. (\ref{b3bis}) is slightly more complicated. Let us start by differentiating $\ln^2(x)\ln^2(1-x)/2$:
\begin{equation}
\frac{1}{2}\frac{d}{dx}\ln^2(x)\ln^2(1-x)=\frac{\ln(x)}{x}\ln^2(1-x)-\ln^2(x)\frac{\ln(1-x)}{1-x}.
\end{equation}
Hence, integrating the latter expression implies
\begin{eqnarray}\label{int1}
&\frac{1}{2}\ln^2(x)\ln^2(1-x)=&\int_0^x\frac{\ln(t)}{t}\ln^2(1-t)\,dt\\& &-\int_0^x\ln^2(t)\frac{\ln(1-t)}{1-t}\,dt.\nonumber
\end{eqnarray}
\begin{widetext}
Now, let us change the variable from $x$ to $-x/(1-x)$ in Eq. (\ref{Lewin762}):
\begin{eqnarray}
\mathrm{Li_4}\left(\frac{-x}{1-x}\right)&=&\ln\left(\frac{x}{1-x}\right)\mathrm{Li_3}\left(\frac{-x}{1-x}\right)-\frac{1}{2}\ln^2\left(\frac{x}{1-x}\right)\mathrm{Li_2}\left(\frac{-x}{1-x}\right)\\
& &+\frac{1}{6}\ln^3\left(\frac{x}{1-x}\right)\ln(1-x)+\frac{1}{6}\int_0^x\ln^3\left(\frac{t}{1-t}\right)\frac{dt}{1-t}.\nonumber
\end{eqnarray}
Expanding the logarithm:
\begin{eqnarray}
\int_0^x\left[\ln^3(t)-3\ln^2(t)\ln(1-t)+3\ln(t)\ln^2(1-t)-\ln^3(1-t)\right]\dfrac{dt}{1-t},
\end{eqnarray}
we notice that, of the four terms of the latter expression, the first and the third have already been dealt with. The fourth is elementary and the second can be evaluated in terms of the others. One finds, after some algebra (see Ref. \cite{Lewin1981}, formula (7.65), p. 204):
\begin{eqnarray}
\dfrac{1}{2}\int_0^x\ln^2(t)\frac{\ln(1-t)}{1-t}\,dt=-\mathrm{Li}_4\left(-\frac{x}{1-x}\right)-\mathrm{Li}_4(x)+\mathrm{Li}_4(1-x)-\mathrm{Li}_4(1),
\end{eqnarray}
and
\begin{eqnarray}\label{int2}
& &\int_0^x \ln^2(t)\frac{\ln(1-t)}{1-t}\,dt
=+2\left[\ln(1-x)\mathrm{Li}_3(x)-\ln(x)\mathrm{Li}_3(1-x)\right]+2\ln(x)\ln(1-x)\mathrm{Li}_2(1-x)\\& & -\frac{\pi^2}{6}\ln^2(1-x)+\frac{1}{12}\ln^2(1-x)\left[6\ln^2(x)+4\ln(x)\ln(1-x)-\ln^2(1-x)\right]+2\mathrm{Li}_3(1)\ln\left(\frac{1}{1-x}\right).\nonumber
\end{eqnarray}
As mention by Lewin, equating Eqs. (\ref{int1}) and (\ref{int2}) simplifies through cancellation of most of the terms due to a particular case of the inversion formula
\begin{equation}
    \mathrm{Li}_n(-x)+(-1)^n\mathrm{Li}_n\left(-\frac{1}{x}\right)=-\frac{1}{n!}\ln^n(x)+2\sum_{r=1}^{\lfloor\frac{n}{2}\rfloor}\frac{\ln^{n-2r}(x)}{(n-2r)!}\mathrm{Li}_{2r}(-1),
\end{equation}
\end{widetext}
where $\lfloor y\rfloor$ denotes the integer part of $y$, and one is left with
\begin{equation}\label{third}
\int_0^{1}\frac{\ln t\ln^2(1-t)}{t}\,dt=-\frac{\zeta(4)}{2}.
\end{equation}
It is worth mentioning that an alternative proof of (\ref{third}) was provided by Connon \cite{Connon2008}. Inserting Eqs. (\ref{first}), (\ref{second}) and (\ref{third}) in the right-hand side of Eq. (\ref{b3bis}) gives the final result
\begin{equation}
b_3=-(-2\zeta(4))-(-6\zeta(4))+2(-\frac{\zeta(4)}{2})=7\zeta(4)=\frac{7\pi^4}{90}.
\end{equation}

\end{document}